\documentclass[preprint2,times]{aastex631}

\usepackage{subfigure}
\usepackage{amsmath}

\usepackage[normalem]{ulem}
\usepackage{soul}
\begin{document}
\title{On the formation height of low-corona and chromospheric channels of the Atmospheric Imaging Assembly (AIA) on board the Solar Dynamics Observatory (SDO)}

\author{Y. Sanjay}
\affiliation{Department of Physics, School of Advanced Engineering\\
University of Petroleum and Energy Studies, Dehradun-248007, Uttarakhand, India}

\author{S.Krishna Prasad}
\affiliation{Aryabhatta Research Institute of Observational Sciences (ARIES) \\
Manora Peak, Nainital-263001, Uttarakhand, India}

\author{R. Erd\'elyi }
\affiliation{ School of Mathematics and Statistics, University of Sheffield\\ 
Hounsfield Road, Sheffield S3 7RH, UK}

\author{ M. B. Kors\'os}
\affiliation{ University of Sheffield, Department of Automatic Control and Systems Engineering,\\ Amy Johnson Building, Portabello Street, Sheffield, S1 3JD, UK}
\author{D. Banerjee}
\affiliation{Aryabhatta Research Institute of Observational Sciences (ARIES) \\
Manora Peak, Nainital-263001, Uttarakhand, India}

\author{P.S. Rawat}
\affiliation{Department of Physics, School of Advanced Engineering\\
University of Petroleum and Energy Studies, Dehradun-248007, Uttarakhand, India}

\begin{abstract}
The multi-wavelength data from the Solar Dynamics Observatory (SDO) is extensively used in studying the physics of the Sun and its atmosphere. In this study, we estimate the formation heights of low-corona and chromospheric channels of the Atmospheric Imaging Assembly (AIA) over the atmospheres of sunspot umbrae during the quiet condition period within 20 different active regions. The upward propagating slow magnetoacoustic waves (slow MAWs) of 3-min period, which are perpetually present in sunspots, are utilized for this purpose. Employing a cross-correlation technique, the most frequent time lag between different channel pairs is measured. By combining this information with the local sound speed obtained from the characteristic formation temperatures of individual channels, we estimate the respective formation heights. The median values of formation heights obtained across all active regions in our sample are 356, 368, 858, 1180, and 1470 km, respectively, for the AIA 1600 {\AA}, 1700 {\AA}, 304 {\AA}, 131 {\AA}, and 171 {\AA} channels. The corresponding ranges in the formation heights are 247 $\--$ 453, 260 $\--$ 468, 575 $\--$ 1155, 709 $\--$ 1937, and 909 $\--$ 2585 km, respectively. These values are measured with respect to the HMI continuum. We find the formation height of UV channels is quite stable (between 250 $\--$ 500 km) and displays only a marginal difference between the AIA 1600 {\AA} and 1700 {\AA} during quiet conditions. On the other hand, the formation height of coronal channels is quite variable.

\end{abstract}


%

\section{Introduction} 
\label{sec:intro}

One of the rapidly expanding realms within solar physics pertains to oscillations in the solar atmosphere, a domain widely explored through data derived from various instruments observing the Sun. The intricate wave behavior observed in the Sun can be ascribed to the density stratification, robust magnetic fields, and non-uniform temperature structure inherent in the solar plasma. The investigation into wave processes is particularly compelling due to their potential role in addressing crucial issues in coronal physics, including solar wind acceleration, CMEs, flare energy release, and the perplexing matter of solar atmospheric heating. 
The predominant periods in sunspot oscillations range from 3 to 5-minute bands \citep[e.g., see review by][]{2015SSRv..190..103J}. According to \citet{1972SoPh...27...80B}, the photospheric oscillations in sunspots typically exhibit notable 5-min periodicities combined with traces of  3-min oscillations, but their power is very small as compared to the surrounding quiet sun.

Interference between sound waves ($p$-modes) that propagate through the Sun's
interior due to chaotic convection causes a dominant five-minute global oscillation on the
solar surface. In higher altitudes, the 5-min oscillation is evanescent as the acoustic cutoff period around the temperature minimum is 280 sec \citep{1983SoPh...87...77R}, and waves
will not be able to travel against gravity above a cut-off period \citep{horace1909theory,1977A&A....55..239B}.  But, 5-minute oscillations have also been observed in higher altitudes \citep[e.g.,][]{2006ApJ...643..540M,10.1093/pasj/59.sp3.S631} because
waves guided by inclined flux tubes have a higher cut-off period and allow for wave
leakage into the solar environment \citep{2004Natur.430..536D,2005ApJ...624L..61D}. But in the umbra, magnetic field lines are mostly
vertical. So, the 5-minute period is the most prevalent in the photosphere of a
sunspot, but as one rises into the chromosphere and corona, the 3-minute period
becomes the dominant period \citep{1987ApJ...312..457T,2016ApJ...817..117S}. The driving mechanism behind these 3-minute oscillations remains a subject of debate. Currently,
magnetoconvection \citep{2017ApJ...836...18C} and external $p$-mode absorption
\citep{2015ApJ...812L..15K} are thought to be the two likely mechanisms causing the
sunspot waves \citep{2015LRSP...12....6K}.

At the photosphere, due to the high plasma-beta ($\beta \approx $100), 
$p$-modes are typically considered acoustic waves \citep{2012PhDT........94O}. However, within the umbra, where 
$\beta \approx $1 \citep{2017ApJ...837L..11C}, these acoustic waves interact with the magnetic field in the magnetized environment of a sunspot, transforming into magneto-acoustic waves (MAWs). As these waves are often compressible, their propagation affects the plasma pressure which makes them easily detectable through intensity oscillations.

According to \citet{2003A&A...403..277R} and \citet{2015A&A...574A.131H}, umbral flashes, originally observed by \citet{1969SoPh....7..351B}, are brightenings with 3 minutes in chromospheric umbrae caused by magneto-acoustic oscillations propagating upward, eventually transforming
into localized shock waves. Propagating slow waves were first detected alongside coronal loops with Transition region and coronal explorer \citep[TRACE;][]{1999SoPh..187..229H} by \citet{1999SoPh..190..249N} \& \citet{2000A&A...355L..23D} and with Solar and Heliospheric Observatory \citep[SoHO;][]{1995SSRv...72...81D} by \citet{1999SoPh..186..207B}. Recently, \citet{2023MNRAS.525.4815R} deduced that the 3-min propagating slow MAWs in coronal loops are instigated by the 3-minute oscillations observed at the footpoints of the fan loops in the photospheric umbrae.  The presence of standing slow MAWs is also found in the solar corona, especially in hot coronal loops \citep[e.g.,][]{2002ApJ...574L.101W, 2021SSRv..217...34W, 2021ApJ...914...81K}. The prevailing consensus is that three-minute period oscillations are the slow MAWs that
travel along magnetic field lines from the photosphere to the corona where they eventually dissipate.

The speeds at which MAWs travel are a combination of Alfv\'en $v_A$ and sound speeds $c_s$. For uniform, static, and infinite background plasma, the fast wave propagates in all directions and has a maximum speed, $v_f$, perpendicular to the magnetic field \citep{2004prma.book.....G} given as

\begin{equation}
\label{eq1}
    v_f = \sqrt{c_s^2 + v_A^2}.
\end{equation}
The slow wave, on the other hand, cannot propagate perpendicular to the magnetic field. Along the magnetic field, it propagates at a tube speed of $v_t$ (for a slender flux tube of radius much less than that of the wavelength of disturbance \citep{1983SoPh...88..179E} given by 

\begin{equation}
\label{eq2}
    v_t = \frac{c_s\ v_A}{\sqrt{c_s^2 + v_A^2}} = \frac{c_s*v_A}{v_f}.
\end{equation}
The plasma $\beta$ is related to $v_A$ and $c_s$ as
\begin{equation}
\label{eq3}
    \beta = \frac{2\ c_s^2}{\gamma \ v_A^2  }.
\end{equation}
Here $\gamma$ is the polytropic index. For regions where the plasma $\beta <\ 1$, Eq.{\,}\ref{eq3} implies $c_s^2 < \ v_A^2$. So, Eq.{\,}\ref{eq1} becomes $v_f \approx v_A$, i.e., the fast MAWs propagate at the Alfv\'en speed. Similarly, Eq.{\,}\ref{eq2} reduces to $v_t \approx \ c_s$, indicating that the slow MAWs propagate at roughly the speed of sound along magnetic field lines. In the absence of a magnetic field, all that exists are simple acoustic waves that propagate isotropically at sound speed. 
In the photospheric umbra, the plasma $\beta \approx \ 1$ \citep{2017ApJ...837L..11C}, and as one ascends higher into the atmosphere, it begins to decrease ($\beta <\ 1$). 
Under coronal conditions, the propagation speed of slow MAWs was measured to be consistent with the projected local sound speed over several coronal loops within active regions \citep[e.g., see review][]{2009SSRv..149...65D}. Therefore, it is safe to assume that, in the umbral atmosphere, slow MAWs predominantly propagate along magnetic field lines with the speed of sound \citep{2002A&A...393..649M}. 

The Helioseismic Magnetic Imager \citep[HMI;][]{2012SoPh..275..207S, 2012SoPh..275..229S} and the Atmospheric Imaging Assembly \citep[AIA;][]{2012SoPh..275...41B} data from the Solar Dynamics Observatory \citep[SDO;][]{2012SoPh..275....3P} provides new avenues for the investigation of oscillations because of their full disk coverage with a high cadence and a high spatial resolution. In this study, we aim to explore the formation height of different SDO/AIA channels within the atmospheres of sunspot umbrae.
\citet{2005ApJ...625..556F} calculated the formation height of TRACE 1700 \r{A} and 1600 \r{A} channels using non-LTE radiation hydrodynamic simulations. These authors find the formation heights of respective channels as  360$\pm$162 km and 430$\pm$92 km. \citet{2012ApJ...746..119R} employed the time lag between different SDO/AIA channels to determine the average wave speed assuming a certain formation height for each channel. Alternatively, \citet{2015ARep...59..959D} applied the cross-correlation technique to determine the average time lags between different SDO/AIA channel pairs and thereby derived the formation height of multiple EUV and UV channels considering the wave propagation speed between the layers as the average local sound speed. Later, \citet{2018SoPh..293....2D} estimated both the height separation and the wave propagation speed between the AIA 1600{\,}{\AA} and 304 \r{A} channels using the projected separation in wave paths and the corresponding time lags. The authors utilize the magnetic field inclination information obtained from the non-linear force-free field extrapolations of photospheric vectors magnetograms from SDO/HMI for this purpose. They assume a constant field inclination between the 1600{\,}{\AA} and 304 \r{A} channels and a fixed formation height of 500 km for the 1600{\,}{\AA} channel in their calculations. Over three different active regions, they obtain a height separation of 500 to 800 km and an average propagation speed of 30{\,}km{\,}s$^{-1}$ between the 1600{\,}{\AA} and 304 \r{A} channels. Notably, there are certain underlying assumptions in every technique.  In the current study, since our objective is to find out the typical formation height of different SDO/AIA channels, we follow an approach similar to \citet{2015ARep...59..959D}, albeit with some important changes, as will be discussed in the subsequent sections. Furthermore, we analyze 20 active regions here as opposed to isolated cases (or only a few) in the previous studies.

In Section{\,}\ref{sec:data}, we describe the details of the data employed, in Section{\,}\ref{sec:analysis} we describe the analysis techniques along with the obtained results, and finally, in Section{\,}\ref{sec:conclusion} we present our summary and conclusions.

\section{Data}\label{sec:data}

We utilize multi-wavelength imaging data acquired by the AIA and HMI telescopes onboard the SDO. Using a robust IDL pipeline developed by Rob Rutten\footnote{https://robrutten.nl/rridl/sdolib/dircontent.html}, the desired subfield data encompassing 20 different active regions were extracted from HMI and AIA. This pipeline derotates and coaligns multi-channel images using low-cadence data from a large disk center subfield. In addition, it applies the necessary roll angle and plate scale corrections to the downloaded level 1.0 data using aia$\_$prep.pro upgrading them to science-grade level 1.5. The final images are rescaled to ensure that each pixel precisely represents 0$\farcs$6 on disk.

The start times of each dataset and the corresponding NOAA AR numbers are listed in Table{\,}\ref{tab:heights}. The duration of each dataset is 1 hour. The extracted field of view is about 180\arcsec$\times$180\arcsec. These data are a subset of the active regions presented in \citet{2018ApJ...868..149K}, where further details are available. 

In this work, we analyze data from 6 wavelength channels, namely 1700{\,}\AA, 1600{\,}\AA, 304{\,}\AA, 131{\,}\AA, and 171{\,}{\AA} from AIA and the continuum channel from HMI. The sunspots within these datasets correspond to different Mount Wilson classes, but no flare was associated with them in the considered period, as identified from the solar monitor site\footnote{https://solarmonitor.org/}. However, eight of the sunspots exhibit the presence of a light bridge (LB; indicated by an asterisk in Table{\,}\ref{tab:heights}) although we exclude the portion of LB in our analysis.

\section{Analysis and results}
\label{sec:analysis}

The near vertical field in a sunspot umbra means it is easier to track a single structure across multiple heights in the solar atmosphere. Therefore, we restrict our analysis to umbral regions of leading sunspot within each active region in our sample. The determination of the umbral region in each dataset is performed on the HMI continuum data following the methodology employed in \cite{2017ApJ...847....5K}. Essentially, this involves calculating a median intensity value over a time-averaged quiet Sun subfield within the HMI continuum data. A fraction of the median value is then used as a threshold to mark the umbra-penumbra boundary. Since LB pixels have higher intensity, selecting a certain \% value will also eliminate the LB sites. The fraction is manually selected until a suitable match with the visual boundary between the umbra, penumbra, and LB pixels is reached.

\begin{figure*}
    \includegraphics[width = 1\linewidth]{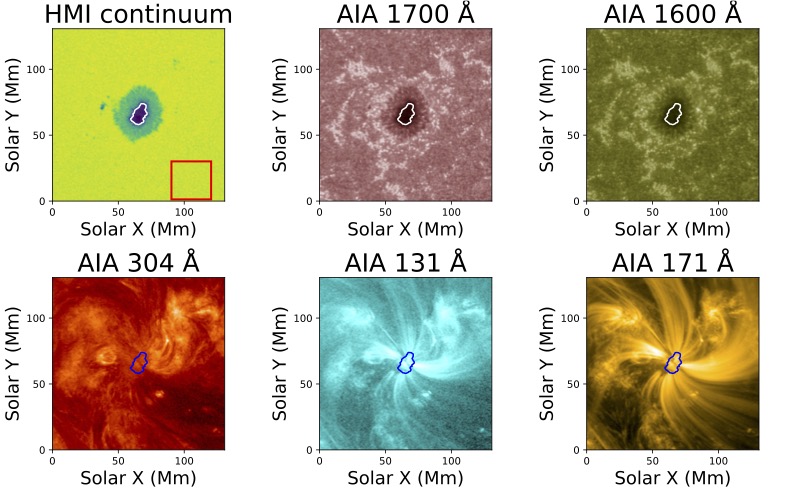}
    \caption{Sample images of a sunspot within AR12553 captured on June 16, 2016, across various wavelengths of SDO channels as shown. The HMI continuum image represents the average intensity over the one-hour duration, while the rest of the images depict the intensity within the first frame. The red box on the HMI continuum image denotes the selected region for calculating the median intensity. The white and blue solid curves delineate the identified boundary between the umbra and penumbra (see text for details).}
    \label{fig:umbra}
\end{figure*}

Owing to the varied characteristics of different active regions in our dataset, we obtain a threshold value of 20-55\% of the median quiet Sun intensity. Fig.{\,}\ref{fig:umbra} illustrates the umbra-penumbra boundary identified over a sample dataset corresponding to AR 12553, which was observed on June 16, 2016. Here, a 25\% of the median intensity value is used. The nearby quiet Sun region identified for computing the median value is also shown in the figure by a red square. The umbral area identified in this manner for all the active regions is listed in Table \ref{tab:heights}.

\begin{figure*}
    \centering
    \includegraphics[width = \linewidth]{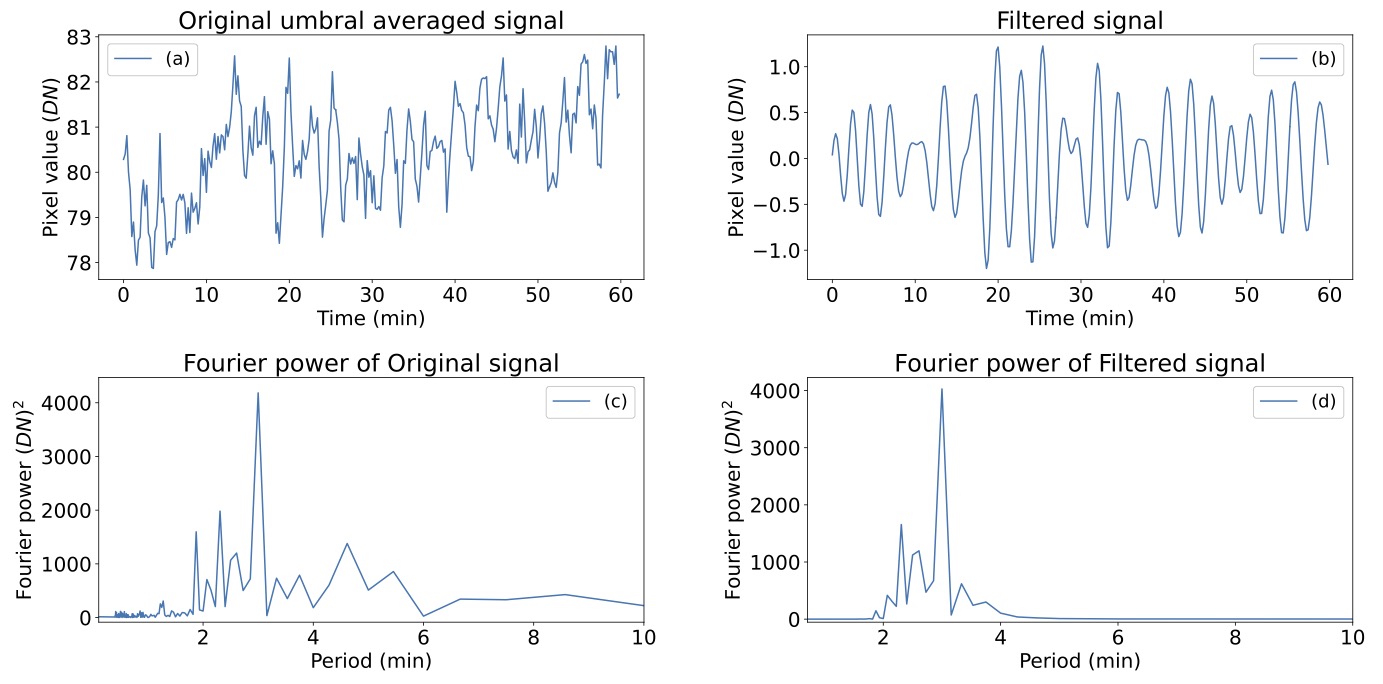}
    \caption{a) Spatially averaged AIA 131 \r{A} intensity from the umbral region of AR 12553 is shown in Fig.{\,}\ref{fig:umbra} b) The signal in a) after passing through a bandpass Butterworth filter retaining oscillations in a 2--4-minute range. c) and d) represent the Fourier power spectra of the original and filtered signals displayed in a) and b), respectively.}
    \label{fig:bwfilter}
\end{figure*}

Given that the 3-minute slow MAWs are predominant in the umbral atmosphere and also generally exhibit a propagating nature, the time series at each umbral pixel within each channel is then Fourier filtered to retain oscillations within a narrow period band of 2-4 minutes. The filtration is accomplished using a third-order Butterworth filter \citep{1977NacEl..31..329K} due to its flat response in the selected pass band. Fig.{\,}\ref{fig:bwfilter} displays a time series along with the corresponding Fourier power spectrum before and after applying this filter. This time series is built from the spatially averaged AIA 131 \r{A} intensity within the umbral region of AR 12553 shown in Fig.{\,}\ref{fig:umbra} where a dominant 3-minute signal can be observed.

The filtered data within different channels are then cross-correlated in pairs to obtain a peak cross-correlation coefficient and the corresponding time lag. For this purpose, we select eight channel pairs, namely, [Continuum, 1700], [Continuum, 1600], [1700, 1600], [1700, 304], [1600, 304], [304, 171], [304, 131],  and [171, 131]. We do not consider all possible combinations here as in \citet{2015ARep...59..959D}, as we reckon that extreme combinations such as [continuum, 171] are prone to large uncertainties. 

Before cross-correlation, the data were linearly interpolated wherever necessary. For instance, when cross-correlating the HMI continuum data with AIA 1700\r{A}, we interpolated the HMI continuum data to match the cadence of AIA 1700\r{A}. But, in cases where the cross-correlation is performed between two channels with the same cadence (e.g.,  EUV channels), no interpolation was done. In each of the channel pairs, the first channel is considered as the reference channel, and the second as the test channel. Considering the dominant vertical field in the umbra leading to a near vertical propagation of slow MAWS, in an ideal scenario, the filtered signal in each pixel in the reference channel can be cross-correlated with the signal from the corresponding pixel in the test channel to estimate the respective wave propagation times. However, it is quite possible that the direction of wave propagation is not aligned precisely with the line of sight, either because the location of the active region is further away from the disk center or because the magnetic field in the umbra inherently possesses some inclination. Both effects could also be in play at the same time. From the location information provided in Table{\,}\ref{tab:heights}, clearly not all active regions are near the disk center. So, this effect is likely not ignorable. 

To alleviate this problem, the signal at each pixel in the reference channel is cross-correlated with the signal from all the pixels within an 11$\times$11 pixel$^2$ box centered over the corresponding pixel (i.e., $\pm$5 pixels in both dimensions), in the test channel. The corresponding time lag values are then obtained from the pixel location that shows the maximum cross-correlation. For a precise estimation of the time lag, the respective cross-correlation curves are fitted with a parabola near the maximum. 
\begin{figure*}
    \centering
    \includegraphics[width = \linewidth]{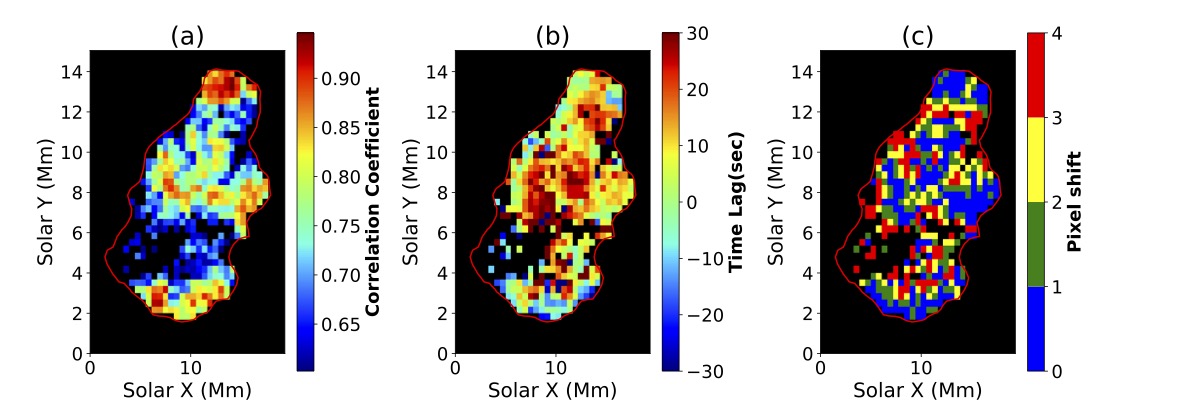}
    \caption{The correlation coefficient (a), time lag (b), and absolute pixel shift (c) maps obtained for AR 12553, considering the AIA 304 and AIA 131 channel pair. Pixel locations with a correlation coefficient below 0.6 are ignored from the time lag and pixel shift calculations. The red curve delineates the boundary between the umbra and penumbra.}
    \label{fig:lags}
\end{figure*}

Fig.{\,}\ref{fig:lags} displays the final cross-correlation coefficient, time lag, and absolute pixel shift maps obtained for AR12553 considering a [304, 131] channel pair. Pixel locations that do not show good correlation (correlation coefficient $\ge$ 0.6) are excluded from our results, as a standard. These observations reveal that the time lag values are not uniform across the umbral region. This spatial non-uniformity can originate from a combination of factors including the complexity of the magnetic field structure and local changes in thermal properties. More particularly, the larger time lag values near the core of the umbra, where a near vertical field is expected, suggest that it is not simply an effect of field inclination.
On the other hand, the dominant blue color in the pixel shift map, corresponding to a shift of 1 pixel or less,  confirms that the field is nearly vertical for most of the region. Note that these values represent the absolute magnitude of the shift. Therefore, there are some locations with fractional values exceeding 1-pixel shift but shown with the same color for simplicity. Nevertheless, the fact that a majority of the pixel locations show a much smaller shift than 5 pixels, with a strong correlation, supports our choice of the box. These results are consistent with \citet{2018SoPh..293....2D} and \citet{2023MNRAS.525.4815R}, who found a maximum shift of 2$\--$3 pixels between various AIA channels.

\begin{figure*}
    
    \subfigure[Side view]{\includegraphics[width = 0.66 \linewidth]{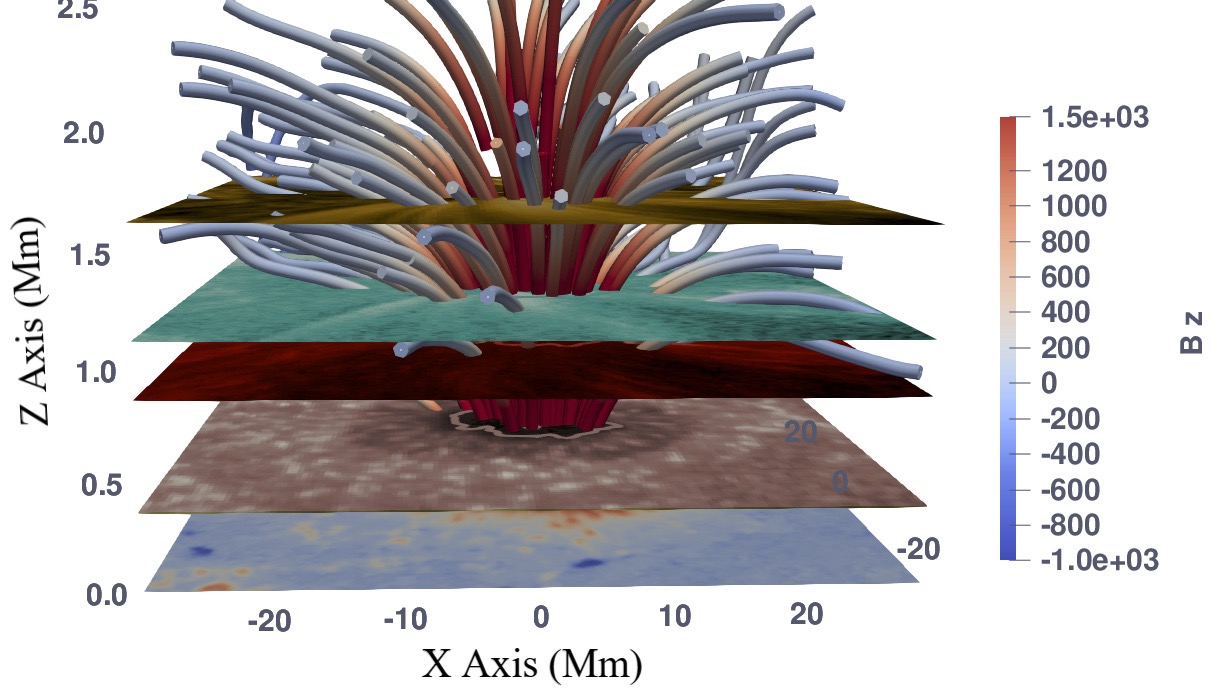}}
    \
    \subfigure[ Top view]{\includegraphics[width = 0.34\linewidth]{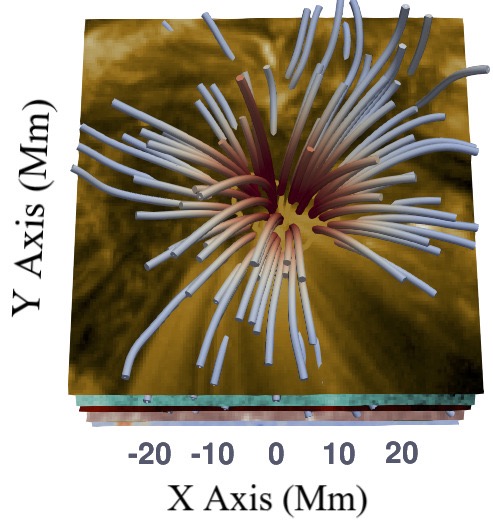}}   
    \caption{Linear Force-free field extrapolations for AR 12186 observed on October 14, 2014. Our focus was on a restricted area around the sunspot to observe the magnetic field lines nearby. All AIA images are stacked according to their formation height obtained from our analysis (see Section{\,}\ref{sec:methods} for details). Note that AIA 1600 and AIA 1700{\,}{\AA} images form very close to each other as compared to other channels so it is difficult to see them distinctly here. a) and b) represent the side view and top view of the extrapolated field lines, respectively. }

    \label{fig:lfff}
\end{figure*}

In order to further verify if the $\pm$5-pixel box is a suitable choice, we perform Linear Force-Free Field (LFFF) extrapolations on two of our active regions using the corresponding HMI SHARP data \citep{2014SoPh..289.3549B}. The specific technique we use employs a fast Fourier transform similar to that pioneered by \citet{1981A&A...100..197A}. The force-free parameter (alpha) is taken from the SHARP data header. Among the two active regions selected, the first is AR 12186, which is quite far from the disk center with position coordinates [166\arcsec, -436\arcsec], and the second is AR 12553, which is relatively near the disk center with position coordinates [11\arcsec, -129\arcsec]. The obtained field extrapolations for AR 12186 are illustrated in Fig.{\,}\ref{fig:lfff}.  Most of the field lines within the umbra are vertical, as expected. To put this quantitatively, we follow each field line within the umbra and estimate the expected horizontal shift in pixels to be on the same field line up to a height of 2.2 Mm. Note that this specific value simply comes from the fact that our height resolution in the $z$-direction is about 0.36 Mm. We find that the maximum shift for AR 12186 is about 5.2 pixels, and the same for AR 12553 is about 4.2 pixels. It may be noted that the locations with large shifts are mostly concentrated near the periphery of the umbra. Besides, as we will find later, the height separation between any of the channel pairs investigated here is much less than 2.2 Mm, so it is likely that the expected shifts are well below 5 pixels at all locations. 

\begin{figure*}
    \centering
    \includegraphics[width = \linewidth]{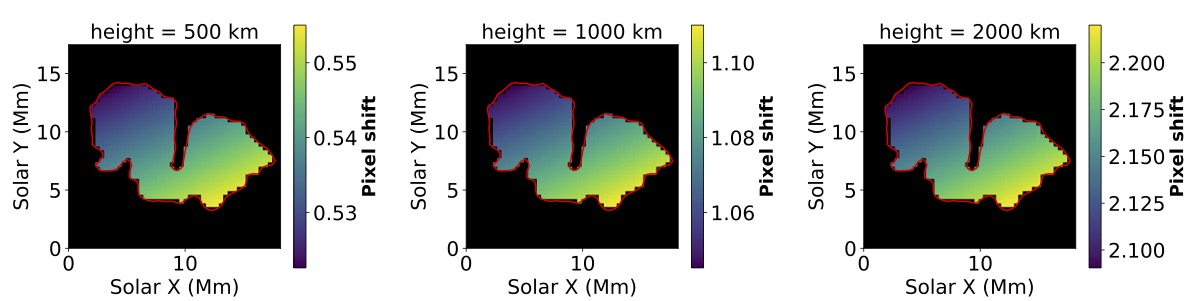}
    \caption{Expected pixel shifts for AR 12186 due to its projection along the line of sight. Different panels show the shifts at different heights as labelled. The red curve represents the boundary between the umbra and the penumbra. The values within the umbra are only displayed.}
    \label{fig:LOS}
\end{figure*}

Another effect that may prevent the wave propagation direction from being radial is the inclination of the local normal with respect to our line of sight. 
This is only an apparent effect and is caused primarily due to the projection of the 3D surface to a 2D plane, normal to our line-of-sight direction. However, because we know the exact 3D coordinates at each pixel location, we can easily calculate the expected shifts as a function of height using simple trigonometric transformations. A pixel shift map, thus obtained for AR 12186, is shown in Fig.{\,}\ref{fig:LOS} for three different heights as labelled. The values within the umbral region are only shown in this figure. The obtained shifts in the southwest region of the umbra are noticeably pronounced. This is because of the position of this active region in the southwest quadrant (see Table \ref{tab:heights}). Regarding the absolute magnitude, at a height of 2 Mm, we obtain a shift of $\approx$2 pixels. This value is on the higher side because AR 12186 is one of the furthest from the disk center in our sample. For instance, at the same 2 Mm height, the corresponding shift obtained for AR 12553 is about 0.7 pixels. Furthermore, if one considers a height separation of 1000 km or less between the selected channel pairs, the expected shifts, even for AR 12186, are going to be of the order of only 1 pixel. In addition, depending on the local orientation of the magnetic field, this shift shall be either added or subtracted. Therefore, the contribution of this effect to the overall shifts can be ignored here.

In summary, our selection of $\pm$5-pixel box in the test channel appears quite reasonable to properly find and trace the wave propagation direction. Note that we use the same size box for all the test channels irrespective of their height. Although the field lines are likely more inclined at greater heights, as we demonstrated above a $\pm$5-pixel box is acceptable for a large height range.

\subsection{Estimating most probable time lag }\label{sec:methods}

The peak correlation coefficient and the corresponding time lag are measured for all pixels within the umbral region between all 8 channel pair combinations selected, as illustrated in Fig.{\,\ref{fig:lags}}. This exercise is applied to all 20 active regions. The obtained time lag values are not uniform across the umbra, as can be seen from Fig.{\,\ref{fig:lags}}. It is possible that this is due to an inherent change in the morphological structure across the umbra, but there is also a likely chance that part of the pixel-to-pixel variation is due to the statistical noise. In either case, we would like to identify a representative time lag value between the layers. This is typically done by taking a simple average across the region \citep{2015ARep...59..959D}, but rather, we opted for the most frequent value here \citep[see for e.g.,][]{2018SoPh..293....2D}. 

To achieve this, we first build histograms of all the time lag values obtained for each channel pair and then fitted these data with a Gaussian curve considering the bin center points. The bin size is considered as 6 s. The peak of the fitted Gaussian is taken as the representative time lag for a particular channel pair. The standard deviation of the Gaussian fit normalized with the square root of the number of bins gives the corresponding uncertainty in the time lag value. 

\begin{figure*}[ht]
    \centering
    \includegraphics[width = \linewidth]{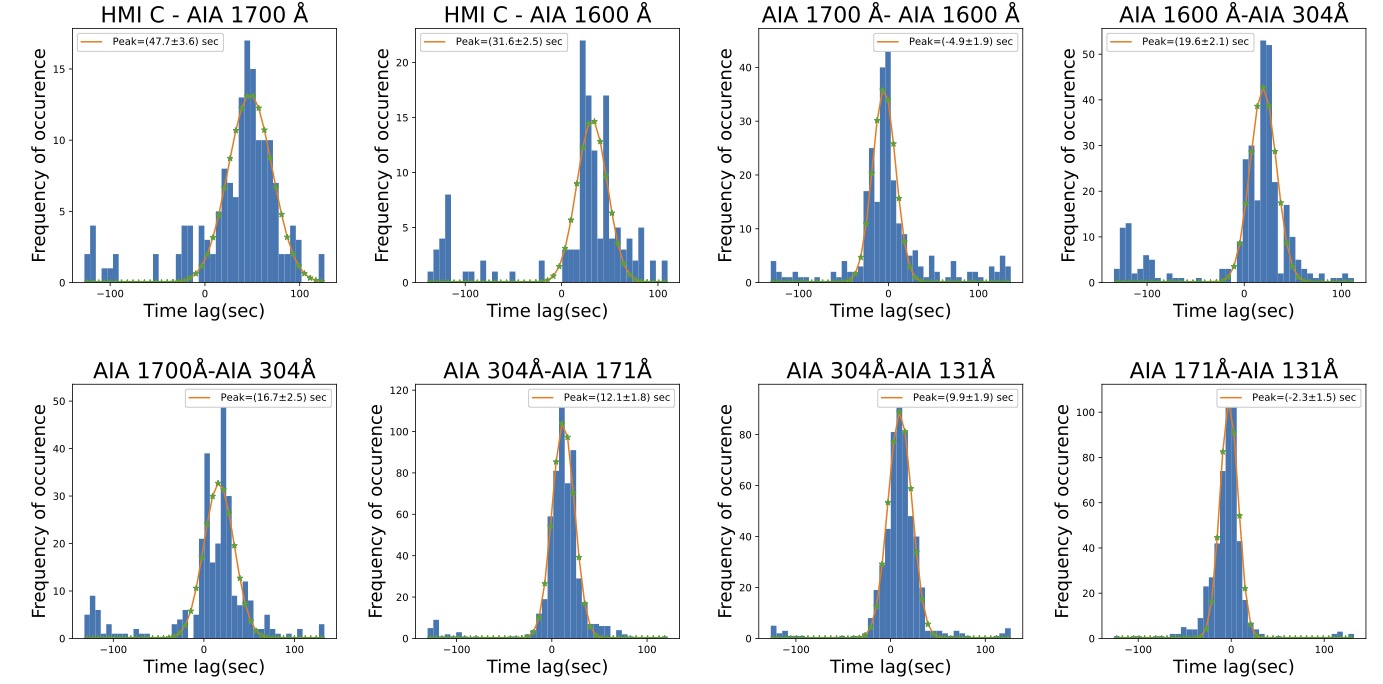}
    \caption{Histograms of time lag values obtained for all 8 channel pairs corresponding to AR 12553 data. A bin size of 6 s is used in these plots. The orange curves represent a Gaussian fit to the histogram data. The corresponding peak values and the respective uncertainties obtained from the fit are listed in the plots.}
    \label{fig:hist}
\end{figure*}

For illustration, the histograms obtained from all 8 channel pairs for AR 12553 are shown in Fig.{\,}\ref{fig:hist}. The Gaussian fits to the data are shown by orange curves, and the respective peak time lag values obtained are listed in the plot.

\subsection{Sound Speed estimation}
\label{sec:speed}
Now that the time lag values are estimated, one could calculate a relative formation height of the test channel with respect to the reference channel, if the propagation speed of the waves is known. Considering the prevailing interpretation of the 3-minute intensity oscillations we observe here as slow MAWs and under the assumption of low plasma-$\beta$ conditions in the umbral atmosphere, the propagation speed can be approximated to the local sound speed. The expression for the sound speed, $c_s$, is given as 
\begin{equation}
    c_s = \sqrt{\frac{\gamma \  k_B \ T}{\mu \ m_H}}.
\end{equation}
Here, $\gamma$ (=5/3) is the polytropic index, $k_B$ is the Boltzmann's constant, $T$ is the temperature, and $\mu \ m_H$ is the mean mass per particle. The $\mu$ value varies with the degree of ionization but a reasonable choice can be 0.61 for corona and 1.25 for the photosphere \citep{1992str..book.....M}. The characteristic temperature for the optically thin EUV channels (AIA 304, 131, and 171) can be obtained from the peak of the temperature response functions. In case of multiple peaks in the response curves, one can choose the peak appropriate for non-flaring plasma conditions. However, this estimation is not as straightforward for the UV channels (AIA 1600 and 1700) because of the optically thick emission. According to \citet{2019ApJ...870..114S}, the active region plage emission in AIA 1700 and 1600 \r{A} is mainly sourced from the continuum created close to the temperature minimum during non-flare times. Indeed, the authors show that the continuum contribution to the overall emission in 1600 \r{A} and 1700 \r{A} channels is about 67 \% and 87 \%, respectively, during quiet periods. So the temperature distinction between them could be marginal. 

On the other hand, during flare periods, AIA 1600 exhibits a dominant contribution from C \textsc{iv}, and AIA 1700 shows a dominant contribution from C \textsc{i}, which highlights a perceptible difference in their formation. Therefore, considering the non-flaring conditions in our data, the characteristic temperatures of AIA 1600 \r{A} and 1700 \r{A} channels can be assumed as $\approx $ 5011 K \citep{2012SoPh..275...17L, 2020ApJ...902...36T}. It may be noted that this is different from the usual consideration of 10$^5$ K as the characteristic temperature for AIA 1600 \r{A} channel \citep[e.g.,][]{2015ARep...59..959D}, but we do not believe such a high value would be applicable here. For HMI Continuum, to determine the characteristic temperature, we assume the emission coming from this layer as a black body radiation. The temperature within the umbra can then be obtained from the intensity ratio e.g., between the umbra and the quiet Sun simply by using a Planck's function as described earlier by \citet{1993A&A...277..639S} and \citet{2017ApJ...837L..11C}. The intensity ratio between the umbra and the quiet Sun ($I/I_{QS}$), can be expressed using Planck's law as

\begin{equation}
\label{eq:planck}
    \frac{I}{I_{QS}} =  \frac{e^{hc/{\lambda k_B T_{QS}}}-1}{e^{hc/{\lambda k_B T}}-1},
\end{equation}
where $h$ is Planck's constant, $c$ is the speed of light, $\lambda$ is the wavelength of HMI continuum (6173 \r{A}), $T_{QS}$ is the temperature of quiet sun (6000 K), $k_B$ is the Boltzmann's constant, and $T$ is the temperature of the umbra. Based on the observed umbral intensity at each pixel an average umbral temperature is obtained. Because the umbral intensity varies from one active region to the other, we obtain different temperatures for different active regions with the overall values ranging from approximately 4176 K to 4764 K.
With these considerations in mind, we estimate characteristic sound speeds for all the 6 wavelength channels as listed in Table \ref{tab:speeds}.

\setlength{\tabcolsep}{10pt}
\begin{table*}
    \centering
      \caption{The primary ion, characteristic temperature, and the sound speed for all 6 channels used in this study considering non-flaring conditions \citep{2020ApJ...902...36T, 2012SoPh..275...17L}. The temperature of the HMI continuum is estimated from the umbral brightness using Planck's function (see text for details). The obtained range across all active regions is listed here.}
    \label{tab:speeds}
    \begin{tabular}{cccc}  
    \hline
         Channel & Primary Ion & Characteristic Temperature & Sound Speed \\
             {}     &   {}         &   (Kelvin)       &   ($\rm km~s^{-1}$)\\
             \hline
         {HMI Continuum} & Continuum & 4176 $\--$4764 & 6.8$\--$7.2\\ 
        {AIA 1700} & Continuum &  5011 & 7.4\\
        {AIA 1600} & Continuum &  5011 & 7.4\\
       {AIA 304} & He II &  50118 & 33.6\\
       {AIA 131} & Fe VIII & 0.45 M & 100.6  \\
       {AIA 171} & Fe IX & 0.79 M &  133.3 \\
       \hline
    \end{tabular}
  
\end{table*}{}

\subsection{Formation Height calculation}

Let us now compute the formation height separation between the two channels by utilizing the average sound speed between each channel pair and the respective time lag. The obtained height separation values for all channel pairs from AR 12553 are listed in Table \ref{tab:lags} along with other relevant parameters. The number of `Good pixels' in this table corresponds to the number of pixel locations with a correlation coefficient $\ge$0.6. The average correlation coefficient across all the good pixels is also listed in the table. Note that we did not take into account the horizontal pixel shifts in converting the time lags to height differences. Although this value is of the order of 1 pixel or less for a majority of the locations (see Fig.{\,}\ref{fig:lags}), it is largely limited by the pixel scale and therefore, prone to accumulate large uncertainties.
\setlength{\tabcolsep}{12pt}
\begin{table*}
    \centering
      \caption{The time lag, average sound speed, and the computed separation in formation height between each channel pair for AR 12553. The values under `good pixels' denote the number of pixel locations with a cross-correlation coefficient of at least 0.6. The total umbral pixels identified for this active region are 637, which is considered the same at all heights. The mean correlation coefficient over the `good pixels' is also listed.}
    \label{tab:lags}
    \begin{tabular}{ccccccccc}
    \hline
        Channel  & Time lag & Average correlation & Average sound & Separation  & Good  \\
         combination & (s) & coefficient & Speed(km{\,}s$^{-1}$) & height (km) & pixels\\
             \hline
        HMI C -1600 &31.6 $\pm$ 2.5&0.65&7.2 & 225$\pm$ 18&132\\
        HMI C- AIA 1700 &   47.7 $\pm$  3.6 & 0.65 &7.2 &339 $\pm $25& 152\\ 
        AIA 1700-1600 &-4.9 $\pm$ 1.9&0.66&7.4&-36$\pm$14&254\\
       AIA 1700-304 &16.7 $\pm$ 2.5&0.67&20.5&342$\pm$52&280\\
       AIA 1600-304 & 19.6 $\pm$ 2.1&0.67&20.5&401$\pm$43&314  \\
       AIA 304-131 & 9.9 $\pm$ 1.9&0.75&67.1&662$\pm$130&507 \\
       AIA 304-171&12.1 $\pm$ 1.8&0.78&83.4&1005$\pm$152&565\\
       AIA 171-131&-2.3 $\pm$ 1.5&0.78&116.9&-264$\pm$174&488\\
       \hline
    \end{tabular}
\end{table*}{}

So far, we found a height separation between each channel pair, but as can be seen from the table, we still cannot assign a particular height to each channel with respect to the photosphere (HMI continuum). This is because the value will change depending on the channel pair chosen. Mathematically speaking, we have more equations than the variables and, therefore, it is an overdetermined problem with no particular solution. In such cases, one can obtain a pseudo solution or, in other words, a solution that is not exact but matches \textit{closely} with all the equations in the least squares sense. This can be achieved by using a singular value decomposition (SVD) method, for example, as followed in \citet{2015ARep...59..959D}.

Essentially, the procedure involves building a set of linear equations from the height separations and solving them using the least square fitting by singular value decomposition method. The set of equations for AR12553 considering the HMI continuum height ($H_c$) as reference are listed below.
\begin{equation}
\label{eq:heights}
\begin{split}
   H_c &= 0 \\
    H_{1600} - H_c &= 225 \\
    H_{1700} - H_c &= 339 \\
    H_{1600} - H_{1700} &= -36 \\
    H_{304} - H_{1700} &= 342 \\
    H_{304} - H_{1600} &= 401 \\
    H_{131} - H_{304} &= 662 \\
    H_{171} - H_{304} &= 1005 \\
    H_{131} - H_{171} &= -264 \\
\end{split}
\end{equation}
This set can be represented as $AX = B$, where $A$ contains the coefficient matrix of each channel's height, $B$ contains the observed height separations (in km), $X$ is what we solve for, which gives the final height (in km) of each channel relative to the HMI continuum. In principle, the solution can be simply obtained from $X = A^{-1}$$\cdot$$B$. 

However, as discussed earlier, the number of independent variables is fewer than the equations, resulting in mismatched dimensions between $A$ and $B$. Hence, using SVD matrix method \citep{book}, we decompose $A$ into $U\cdot$$S\cdot$$V^T$, where $U$ is the matrix of the orthonormal eigenvectors of $A\cdot$$A^T$, $V^T$ is the transpose of matrix containing the orthonormal eigenvectors of $A^T\cdot$$A$, and $S$ is the diagonal matrix with elements equal to the root of eigenvalues of $A\cdot$$A^T$ or $A^T\cdot$$A$. Thus, the solution for $X$ can be written as  $X=V$$\cdot$$S^{-1}$$\cdot$$U^T$$\cdot$$B$. 

Furthermore, in order to incorporate the uncertainties we obtained on the height separations, instead of solving $AX = B$, we solve $A^\prime$$X = B^\prime$, where $A^\prime$$ = w\cdot$$A$ and $B^\prime$ $= w\cdot$$B$ with $w$ as the matrix that contains the reciprocals of the respective errors. 

\begin{figure*}[ht!]
    \centering
    \includegraphics[width = \linewidth]{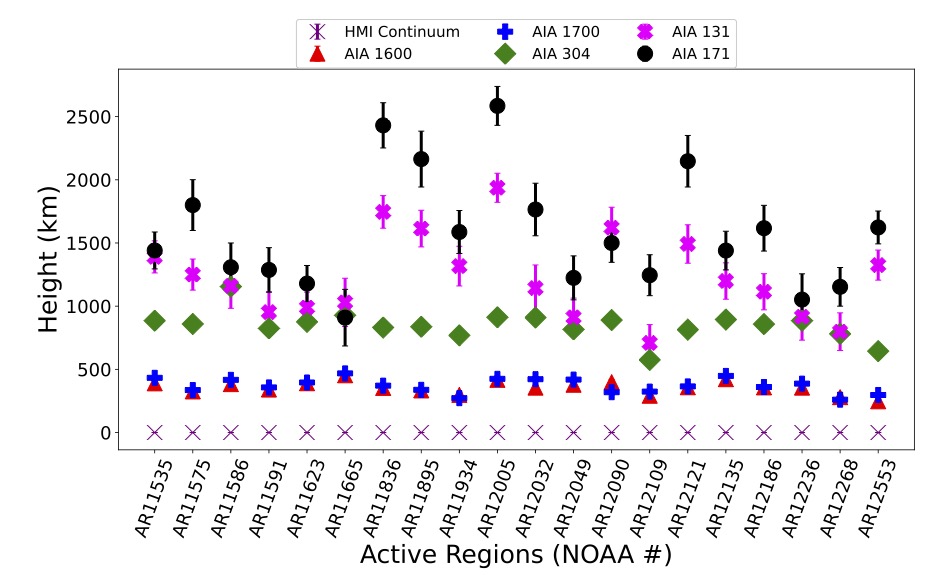}
    \caption{Formation heights of different SDO/AIA channels with respect to HMI Continuum. The vertical bars denote the corresponding uncertainties. Different colors/symbols represent the data for different channels as listed. The uncertainties for AIA 1600, 1700, and 304 \r{A} cases are smaller than the symbol size and, therefore, not clearly visible in the plot. Notably, AIA 1700 \r{A} can be seen to form above AIA 1600 \r{A} in a majority of cases. The AIA 131 \r{A} channel appears to form consistently between AIA 304 \r{A} and AIA 171 \r{A} channels except in two cases.}
    \label{fig:heights}
\end{figure*}

Finally, the solution in $X$ will have an uncertainty given by the covariance matrix, $C$, expressed as $C = V$$\cdot$$S^{-2}$$\cdot$$V^T$. The formation heights thus obtained for AR 12553 are 247$\pm$15 km, 297$\pm$17 km, 644$\pm$36 km, 1325$\pm$119 km, and 1623$\pm$130 km, respectively, for the AIA 1600, 1700, 304, 131, and 171 channels with respect to the HMI continuum. Similarly, the formation heights obtained for all 20 active regions are listed in Table \ref{tab:heights}. Additionally, Fig.{\,}\ref{fig:heights} shows a graphical representation of these data.

\setlength{\tabcolsep}{3pt}
\begin{table*}
    \centering
     \caption{The computed formation heights of UV and EUV channels with respect to the HMI continuum for all 20 active regions. The start times of the datasets, the respective NOAA active region numbers, the umbral area, and the umbral average magnetic field are also listed. The asterisk symbols next to the active region numbers denote the cases where a LB is present.}
  
    \label{tab:heights}
    \begin{tabular}{cccccccccc}
    
    \hline
        Start time&Location   & NOAA & Umbral&Avg. abs.  & AIA 1600 & AIA 1700 & AIA 304 & AIA 131 & AIA 171  \\
         (UT)&(\arcsec) & Number & area  & Mag.&(km) &(km)&(km) & (km)&(km)\\
         &&&$(Mm)^2$&field(G)&&&&\\
             \hline
             
         2012 Aug 5 12:00 &(172,221)&11535&48&1534&$389\pm 20$&$433\pm 19$&$885\pm 21$&$1390\pm 127$&$1441\pm 146$\\
         2012 Sep 23  12:00&(- 215,15)&11575*&94&1294&$325\pm 17$&$336\pm17$&$859\pm 26$&$1250\pm 123 $&$1800\pm 201$\\
        2012 Oct 12 13:00 & (- 83,- 303) &11586&41&1379&$383\pm16$&$417\pm 16$&$1155\pm 48$&$1160\pm 177$&$1308\pm 192$\\
        2012 Oct 19 12:00&(240,30)& 11591&100&1525&$340\pm 19$ &$357\pm 19$&$824\pm 22$&$953\pm 160$&$1287\pm 176$ \\
        2012 Dec 4  12:00&(230,116)&11623*&78&1474&$390\pm 24$&$396\pm 23$&$876\pm 27$&$989\pm 122 $&$1179\pm 143$\\
        2013 Feb 5 12:00 & (355,289)&11665&78&1447&$453\pm 16$&$468\pm 15$&$928\pm 23$&$1030\pm 191$&$909\pm 224$\\
        2013 Sep 3  12:00&(276,70)&11836*&50&1400&$353\pm 23$&$371\pm 22$&$831\pm 24$&$1746\pm 129 $&$2431\pm 179$\\
        2013 Nov 14 13:00 &(- 278,-316)&11895&56&1339&$333\pm 16$&$338\pm 16$&$837\pm 20$&$1614\pm 145$&$2164\pm 221$\\
        2013 Dec 28 13:30 &(576,-215)&11934&53&1282& $298\pm 16$&$272\pm 16$&$769\pm 19$&$1317\pm 157$&$1587\pm 170$\\
       2014 Mar 18 12:00&(13,317)&12005&100&1529& $417\pm 23$&$425\pm 23$&$912\pm 26$&$1937\pm 115$&$2585\pm 154$\\
        2014 Apr 13 12:00 &(-68,280)&12032&88&1497 &$355\pm 19$&$422\pm 19$&$910\pm 30$&$1143\pm 183$&$1765\pm 208$\\
        2014 May 3 12:00&(48,-80)&12049*&93&1571&$378\pm22$&$419\pm22$&$816\pm 39$&$912\pm 152 $&$1224\pm 174$\\
        2014 Jun 16 03:50 &(-196,380)&12090&62&1374&$399\pm 21$&$319\pm 19$&$890\pm 26$&$1621\pm 162$&$1500\pm 153$\\
        2014 July 7  12:00&(-209,-193)&12109*&289&1572&$291\pm 22$&$324\pm 23$&$575\pm 40$&$709\pm 145 $&$1245\pm 162$\\
        2014 July 28  12:00&(87,40)&12121*&57&1457&$358\pm 15$&$365\pm 16$&$814\pm 21$&$1492\pm 152 $&$2147\pm 204$\\
        
        2014 Aug 11 12:00&(33,132)& 12135&86&1441&$422\pm 25$&$448\pm 22$&$894\pm 29$&$1199\pm 144$&$1440\pm 153$\\
        2014 Oct 14 12:00 &(166,-436)&12186&93&1282&$357\pm 24$&$360\pm 24$&$858\pm 28$&$1114\pm 143$&$1617\pm 181$\\
        2014 Dec 17 12:00 &(-35,492)&12236*&38&1285&$355\pm 14$&$387\pm 13$&$885\pm 16$&$916\pm 185$&$1052\pm 204$\\
        
        2015 Jan 29 10:30 &(347,330)&12268*&139&1535&$280\pm 21$&$260\pm 21$&$781\pm 26$&$798\pm 149$&$1153\pm 153$\\
        
        2016 Jun 16 11:00&(11,-129)& 12553 &123&1774&$247 \pm 15$ &$297\pm 17$ &$644\pm 36$ & $1325\pm 119$ &$1623\pm 130 $\\

       \hline
    \end{tabular}
   
\end{table*}

\section{Discussion and Conclusions}
\label{sec:conclusion}

Utilizing the time lags between propagating 3-minute oscillations within the sunspot umbrae, we estimated the formation heights of 5 AIA channels, 1600{\,}\AA, 1700{\,}\AA, 304{\,}\AA, 131{\,}\AA, and 171{\,}\AA, with respect to HMI continuum over 20 different active regions (see Fig.{\,}\ref{fig:heights} and Table \ref{tab:heights}). The median values of the obtained formation heights for these channels are 356$_{247}^{453}$, 368$_{260}^{468}$, 858$_{575}^{1155}$, 1180$_{709}^{1937}$, and 1470$_{909}^{2585}$ km, respectively. The subscript and superscript values here indicate the overall range in heights obtained across all the active regions studied. According to \citet{2006ASPC..358..193N}, the continuum near Fe{\,}\textsc{i} 6173 \r{A} forms at an altitude of 21 km in the umbral atmosphere. If one takes this into account, all our formation heights shall be shifted accordingly.

Among the UV channels, it is observed that in about 17 cases, the 1700{\,}{\AA} channel forms above 1600{\,}{\AA} while it is the other way round in the rest. The 304{\,}{\AA} channel forms above both the UV channels, followed by the 131{\,}{\AA} channel, in all cases. The 171{\,}{\AA} channel forms higher than the 131{\,}{\AA} in all cases except two (AR 11665 \& AR 12090). However, the height separation between the channels in those two cases is marginal. For AR 11665 \& AR 12236, the formation heights of all three EUV channels are very close to each other. Also, as can be seen from Fig.{\,}\ref{fig:heights}, the variation in formation heights of coronal channels (131{\,}{\AA} \& 171{\,}{\AA}) is substantial in comparison with that from the lower atmospheric channels.

Some of our active regions contain sunspots with LBs, as noted before. \citet{1997ApJ...484..900L} reported that LBs can cause magnetic field disruptions. When compared to the surrounding umbra, they found that LBs had more horizontal fields. More recently, \citet{2020ApJ...893L...2F} suggested that magnetoconvection could cause LBs and penumbrae to be split at the photosphere, suggesting that both could share the same magnetic source. According to \citet{2018A&A...609A..73R}, LBs encompass multi-thermal structures extending up to the transition region (TR). Moreover, the chromospheric temperature, as well as AIA 171, 211 Å intensity, are shown to be enhanced in the LBs \citep{2023ApJ...942...62L}.  

In light of these results, we propose that the magnetic structure within the LBs is significantly different from that in the umbra. Therefore, we strictly excluded the pixel locations of LBs from our analysis. Perhaps as a consequence of this selection, we do not observe any peculiar behaviour in the formation heights of AIA channels, even in the case of active regions with LBs. 

In Table \ref{tab:heights}, we also list the umbral area and the average absolute magnetic field strength of the sunspots within individual active regions. The data of the umbral area are a bit discontinuous with a large gap between $\approx 150–280 (Mm)^2$, but within this sample, we did not find any systematic dependence of formation height on the umbral area. The magnetic field strength, too, did not appear to show any influence on the formation height, but perhaps our sample is not large enough to show a consistent behavior.

In general, our formation height values agree well with the results reported in the literature. Employing 1D radiation hydrodynamic simulations, \citet{2005ApJ...625..556F} obtained a formation height of 360$\pm$162 km and 430$\pm$92 km for the TRACE 1700{\,}{\AA} and 1600{\,}{\AA} channels, respectively. Using RATAN-600 multiwavelength radio observations, \citet{Kaltman2012TheAS} find the altitudes of the base of the transition region (10000 K plasma) and the corona (2.5 MK plasma) as 500 and 2000 km, respectively.  \citet{2015ARep...59..959D,2018SoPh..293....2D} observe that the separation between AIA 1600 \r{A} and 304 \r{A} channels lies in the range of 500$\--$800 km. More recently, \citet{2022FrASS...9.1118D} shows that the transition region above a solar umbra with a temperature of $\approx$ 131,440 K (analogous to our 304{\,}{\AA} channel) is situated at 1080$\pm$20 km. Our results for the respective channels fall in the same range of heights. 

However, if one looks at the order of formation, most previous studies consider the 1600{\,}{\AA} channel to be forming above the 1700{\,}{\AA} channel in agreement with the results of \citet{2005ApJ...625..556F}. But, as we find here, this is not the case for 17 out of 20 active regions. As discussed in Section \ref{sec:speed}, this discrepancy is mainly due to the assumption that 1600{\,}{\AA} channel has a dominant contribution from C \textsc{iv}, which is only true during flare periods \citep{2019ApJ...870..114S}. Under quiet atmospheric conditions that are applicable here, the continuum contribution is rather dominant, resulting in marginal differences in height with respect to the 1700{\,}{\AA} channel, as is the case found here. One could also argue that the negative time lags found between the [1700, 1600] channel pair (see Table \ref{tab:lags}) actually implies a downward propagation of the wave. 

Indeed, numerous studies in the literature suggest both upward and downward propagating waves, especially in the lower solar atmosphere \citep[see the reviews by][]{2015LRSP...12....6K, 2023LRSP...20....1J}. While this is certainly a possibility if we only consider the [1700, 1600] channel pair, the results from [continuum, 1600] and [continuum, 1700], when considered in combination, would not be compatible with this scenario. Therefore, we are confident that our results indicate a dominant upward propagation of slow MAWs. Nevertheless, if one is interested in oscillations at individual pixels,  we agree that there may be cases representing both upward and downward propagating waves (see Fig. \ref{fig:hist}).

Lastly, we would like to note some of the important caveats of these results. i) The formation heights estimated here are limited to the umbral atmosphere under quiet conditions. These values may differ for other atmospheric structures, including the surrounding penumbra. ii) Also, the presence of any sort of activity in the region can impact the results. iii) Furthermore, we consider a single formation height for the entire umbra, albeit in a particular wavelength channel. 
Given the dynamic and complex nature of the solar environment, this assumption may not be always valid. iv) In addition, the propagation speed of slow MAWs is assumed to be the local sound speed in each layer which in turn is dependent on the characteristic temperature of the respective wavelength channel. While this is valid for low plasma $\beta$ conditions, which are typical for an umbra, any deviations, especially in the periphery of the umbra, can affect the overall results. 

We hope that future follow-up studies are inspired here that will improve upon our results by addressing these issues.

\section*{acknowledgements}
The data used here are courtesy of NASA/SDO and the HMI and AIA science teams. SKP is grateful to SERB for a startup research grant (No. SRG/2023/002623).
MBK acknowledges support by UNKP-22-4-II-ELTE-186, ELTE Hungary and is grateful for the Leverhulme Trust Found ECF-2023-271. RE is grateful to STFC (UK, grant number ST/M000826/1) and PIFI (China, grant number No. 2024PVA0043). RE and MBK also thank for the support received from NKFIH OTKA (Hungary, grant No. K142987). This work was also supported by the NKFIH Excellence Grant TKP2021-NKTA-64. 


\end{document}